\documentstyle[11pt,newpasp,epsf,twoside]{article}
\begin{document}

\addcontentsline{toc}{section}{
Kinematics of Molecular Hydrogen Emission from \\ Planetary and
Pre-planetary Nebulae \\
\hspace{.25in} {\it J.H. Kastner, I. Gatley, \& D.A. Weintraub}}

\title{Kinematics of Molecular Hydrogen Emission from Planetary and
Pre-planetary Nebulae}
\author{Joel H. Kastner, Ian Gatley}
\affil{Chester F. Carlson Center for Imaging Science, Rochester Institute of
Technology, 54 Lomb Memorial Dr., Rochester, NY 14623}
\author{David A. Weintraub}
\affil{Dept.\ of Physics \& Astronomy, Vanderbilt University, Nashville, TN}

\begin{abstract}

We report results from a program of high-resolution
spectral mapping of rotational H$_2$ emission
from bipolar planetary and pre-planetary nebulae. Long-slit
spectra obtained with the NOAO Phoenix near-infrared spectrometer
allow us to probe the molecular kinematics of these nebulae
at moderate spatial resolution. We find strong evidence of a
component of rotation in the equatorial H$_2$ emission from
the Egg nebula (RAFGL 2688). In this nebula and in the
pre-planetary nebula RAFGL 618, the H$_2$ kinematics point to
the recent emergence of high-velocity polar flows,
which likely mark the fairly sudden terminations of the red giant
phases of their central stars. The classical bipolar
planetary NGC 2346 displays distinct kinematic
components, which we interpret as arising in the
morphologically distinct equatorial and polar regions of the nebula. 
The H$_2$ rings observed in the Phoenix position-velocity
maps of this nebula support the hypothesis that ring-like
planetaries that display H$_2$ emission possess bipolar structure.

\end{abstract}


\section{Introduction}

The presence of molecular hydrogen
emission is now recognized as a reliable indicator of
bipolar structure in planetary nebulae (Zuckerman \& Gatley
1988; Kastner et al.\ 1994, 1996). While the polar lobes
often display H$_2$, the molecular emission is, with few
exceptions, brightest toward the waists of bipolar
planetaries. These molecule-rich regions of planetary
nebulae (PNs) appear to be the remnants of circumstellar
disks or tori formed during previous, asymptotic giant
branch (AGB) or post-AGB phases of the central
stars. Furthermore, the available evidence suggests that the
onset of H$_2$ emission postdates the AGB stage but precedes
the formation of the PN (Weintraub et al.\ 1998). This onset
likely signals the beginning of a
high-velocity, collimated, post-AGB wind, which shocks the
previously ejected, ``slow,'' AGB wind and thereby produces
the observed H$_2$ emission (Kastner et al.\ 1999).

These observations make clear that further investigations of
H$_2$ emission are important to our understanding of the
origin of bipolarity in PNs. It is of particular interest to
establish whether the spatially distinct waist and lobe
H$_2$ emission regions are kinematically distinct as well
and, furthermore, whether the kinematics bear evidence of
the presence of circumstellar disks and/or high-velocity
polar flows. To this end, we have undertaken a program of
spectroscopic mapping of near-infrared H$_2$ emission from planetary and
pre-planetary nebulae at high spectral resolution. First
results from this program were presented in Weintraub et
al.\ (1998), in which H$_2$ emission was detected from a
pair of bipolar pre-planetary nebulae (PPNs), and in Kastner
et al.\ (1999), where we described preliminary results for
the seminal PPN RAFGL 2688. Here we present further analysis
and interpretation of H$_2$ velocity mapping of RAFGL 2688,
as well as H$_2$ velocity mapping results for the PPN RAFGL
618 and the bipolar planetary nebula NGC 2346 (see also Arias \& Rosado,
in this volume).

\section{Observations}

Data presented here were obtained with the
NOAO\footnote{National Optical Astronomy Observatories is
operated by Associated Universities for Research in
Astronomy, Inc., for the National Science Foundation.}
Phoenix spectrometer on the 2.1 m telescope at
Kitt Peak, AZ, in 1997 June (RAFGL 2688) and 1997 December (RAFGL
618, NGC 2346). Phoenix illuminates a $256\times1024$ section of an Aladdin 
InSb detector array. The spectrograph slit was $\sim60''\times1.4''$
oriented approximately east-west. The velocity resolution was
$\sim4$ km s$^{-1}$ and the spatial resolution $\sim1.5''$ 
at the time these spectra were obtained. 
A spectral image centered near the 2.121831 $\mu$m $S(1)$, $v=1-0$
transition of H$_2$ was obtained at each of 12 spatial positions as the slit
was stepped from south to north across RAFGL 2688. The step size, $1.0''$,
provided coverage of the entire H$_2$ emitting region with spatial
sampling approximating the slit height. For RAFGL 618, whose bright H$_2$
emission regions are oriented almost perfectly east-west (Latter et al.\
1995), parallel to the Phoenix slit, we obtained a single spectral image
centered on the object. For NGC 2346 we obtained spectral images at selected 
positions near the waist of the nebula.
Spectral images were reduced and wavelength calibrated as
described in Weintraub et al.\ (1998). For the RAFGL 2688 data,
the reduced spectral images were stacked in declination according to the
commanded telescope offsets, to produce a (RA, dec, velocity) data cube of
H$_2$ emission.

\section{Results and Discussion}

\begin{figure}[htbp]

\plotone{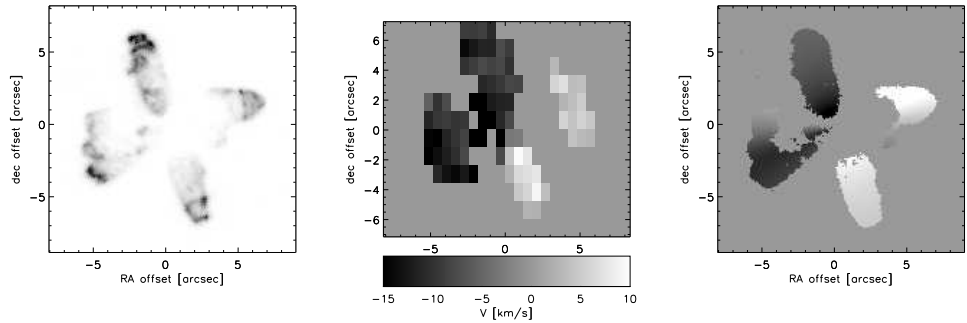}

\caption{Comparison of the model and observed 
H$_2$ velocity fields of RAFGL 2688. The velocity greyscale bar applies to
the center and right panels.
The HST/NICMOS H$_2$ image (Sahai et al.\ 1998) is shown at left. 
The observed velocity field (center) consists of velocity centroids
calculated from the Phoenix data cube. 
In the model (right), we set the equatorial expansion velocity at $v_e=5$ km
s$^{-1}$ and the equatorial rotation velocity at $v_r=10$ km s$^{-1}$.
The comparison indicates that there is reasonable qualitative and
quantitative agreement between model and data for this
choice of parameters. }

\end{figure}

\subsection{RAFGL 2688}

Kastner et al.\ (1999) presented selected velocity planes
from the RAFGL 2688 Phoenix data cube. The four principal
``lobes'' of H$_2$ emission seen in direct H$_2$ images
(e.g., Sahai et al.\ 1998) are also apparent in these Phoenix
data, with one pair oriented parallel to the polar axis
(roughly N-S) and one perpendicular (roughly E-W).  Each of
these H$_2$ lobe pairs displays a velocity gradient, with
the N and E lobes blueshifted by up to $\sim30$ km s$^{-1}$
and the S and W lobes similarly redshifted. However, the N-S
and E-W H$_2$ lobe pairs differ in their detailed kinematic
signatures (Kastner et al.). 

The H$_2$ kinematic data for RAFGL 2688, like velocity maps
obtained from radio molecular line emission, can be
described in terms of a multipolar system of purely radially
directed jets (Cox et al.\ 1997; Lucas et al., these
proceedings).  Given the constraints imposed by Phoenix and
{\it Hubble Space Telescope} data,
however, this model would require that the ``equatorial''
components located east and west of the central star are in
fact directed well above and below the equatorial plane,
respectively (Kastner et al.\ 1999).  If one postulates
instead that the E-W H$_2$ emission lobes are confined to
the equatorial plane of the system --- a hypothesis that
appears to be dictated by certain details of the H$_2$
surface brightness distribution, as well as by simple
symmetry arguments --- then one must invoke a model
combining radial expansion with a component of azimuthal
(rotational) velocity along the equatorial plane (Kastner et
al.). In a forthcoming paper we will compare these two
alternative models in more detail.  Here, we describe a
specific formulation of the latter (expansion $+$ rotation)
model that reproduces many of the salient features of the
Phoenix data.

To construct this empirical model of the H$_2$ kinematics of RAFGL 2688, we are
guided by the basic results described above. That is, the
polar lobes are characterized by velocity gradients in which the fastest
moving material is found closest to the star, and the slowest moving
material is found at the tips of the H$_2$ emission regions. For simplicity,
we assume this behavior can be described by an inverse power law relationship 
between velocity and radius. For the
equatorial plane H$_2$ emission, meanwhile, we assume a
combination of azimuthal (rotation) and radial (expansion) velocity
components, whose magnitudes we denote by $v_r$ and $v_e$, respectively. 

To constrain these model parameters, we compared model velocity field images
with a velocity centroid image which we obtained
from the Phoenix data cube. For the polar  lobes, we find that 
the exponent of the inverse power law velocity-distance relationship 
is roughly $\sim0.7$ and that the outflow
velocities at the tips of the N and S lobes are $\sim20$ km s$^{-1}$. 
For the equatorial regions, good agreement between model and data is obtained
for values of $v_e$ and $v_r$
that lie in the range $5-10$ km s$^{-1}$, with the additional constraint $v_e
+ v_r \sim 15$ km s$^{-1}$. An example of the results for a representative
model (with $v_e = 5$ km s$^{-1}$ and $v_r = 10$ km s$^{-1}$) 
is displayed in Fig.\ 1. There is clear qualitative agreement
between the model and observed velocity images for these parameter values,
in the sense that the overall distribution of redshifted and blueshifted
emission is captured by the model. Furthermore, this model reproduces 
specific details of the observed H$_2$ velocity distribution, such as
the approximate magnitudes and positions of the velocity extrema in the 
four H$_2$ lobes. While this model is by no means unique, the
comparison of calculated and observed velocity fields provides further
support for a component of azimuthal velocity along the equatorial plane of
RAFGL 2688, and offers an indication of the magnitude of this 
``rotational'' component relative to the components of radial expansion both 
parallel and perpendicular to the polar axis of the system.

\subsection{RAFGL 618}

\begin{figure}[ht]

\plotone{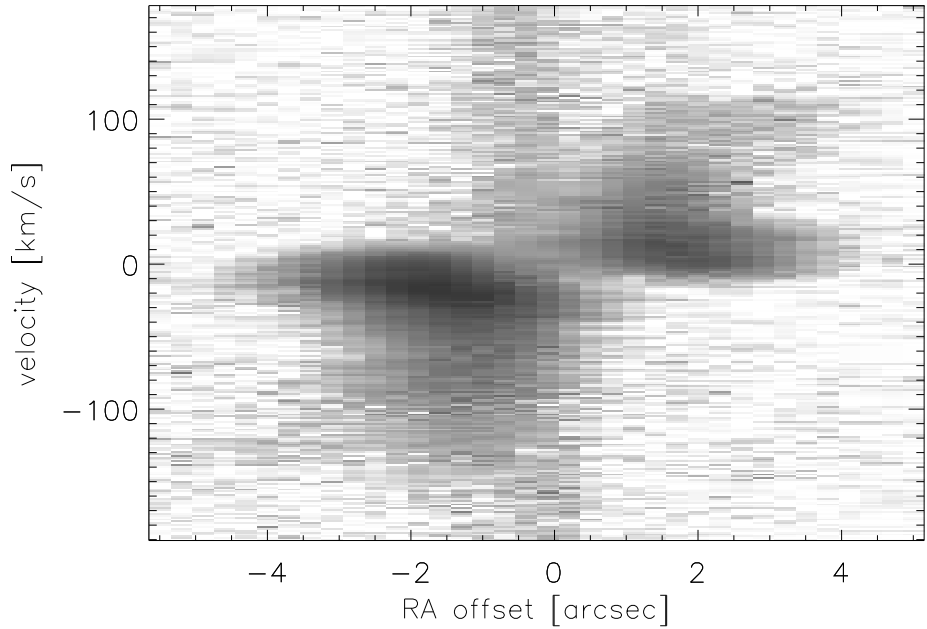}
\caption{Phoenix spectral image of RAFGL 618. The velocity scale of the
image is centered on the systemic velocity of RAFGL 618.
East is to the left. The image is displayed in a logarithmic
greyscale to bring out the line wing emission, which extends to at
least $\sim \pm 100$ km s$^{-1}$. The vertical band across
the image at RA offset $\sim0''$ is produced by continuum
emission from the vicinity of the central star.}

\end{figure}

The Phoenix spectral image obtained for RAFGL 618 is displayed in Fig.\ 2. 
Bright H$_2$ emission is detected along the entire polar axis of RAFGL 618. 
These data demonstrate further that very high velocity H$_2$ emission 
is present in this bipolar outflow. 
The highest velocity molecular material is found closest to the central star
of RAFGL 618. We conclude that the velocity gradients along the polar axes 
of both RAFGL 2688 and RAFGL 618 trace rapid transitions from the ``slow,''
spherically symmetric winds of their AGB progenitors to faster, collimated,
post-AGB winds (Kastner et al.\ 1999). 

\subsection{NGC 2346}

\begin{figure}[ht]

\plotone{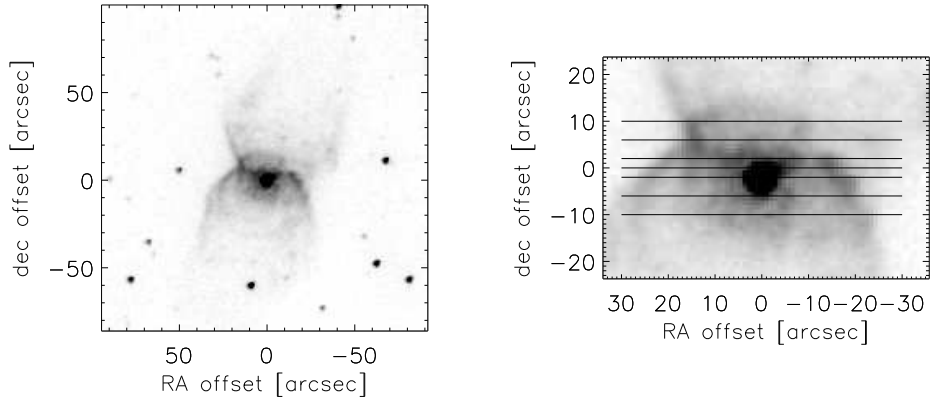}

\caption{Left: Image of NGC 2346 in the 2.12 $\mu$m line of H$_2$
obtained with the NOAO Cryogenic Optical Bench (COB; Kastner
et al.\ 1996). Right: Central region of the COB image,
illustrating the slit positions used for Phoenix observations.}

\end{figure}

\begin{figure}[ht]

\plotone{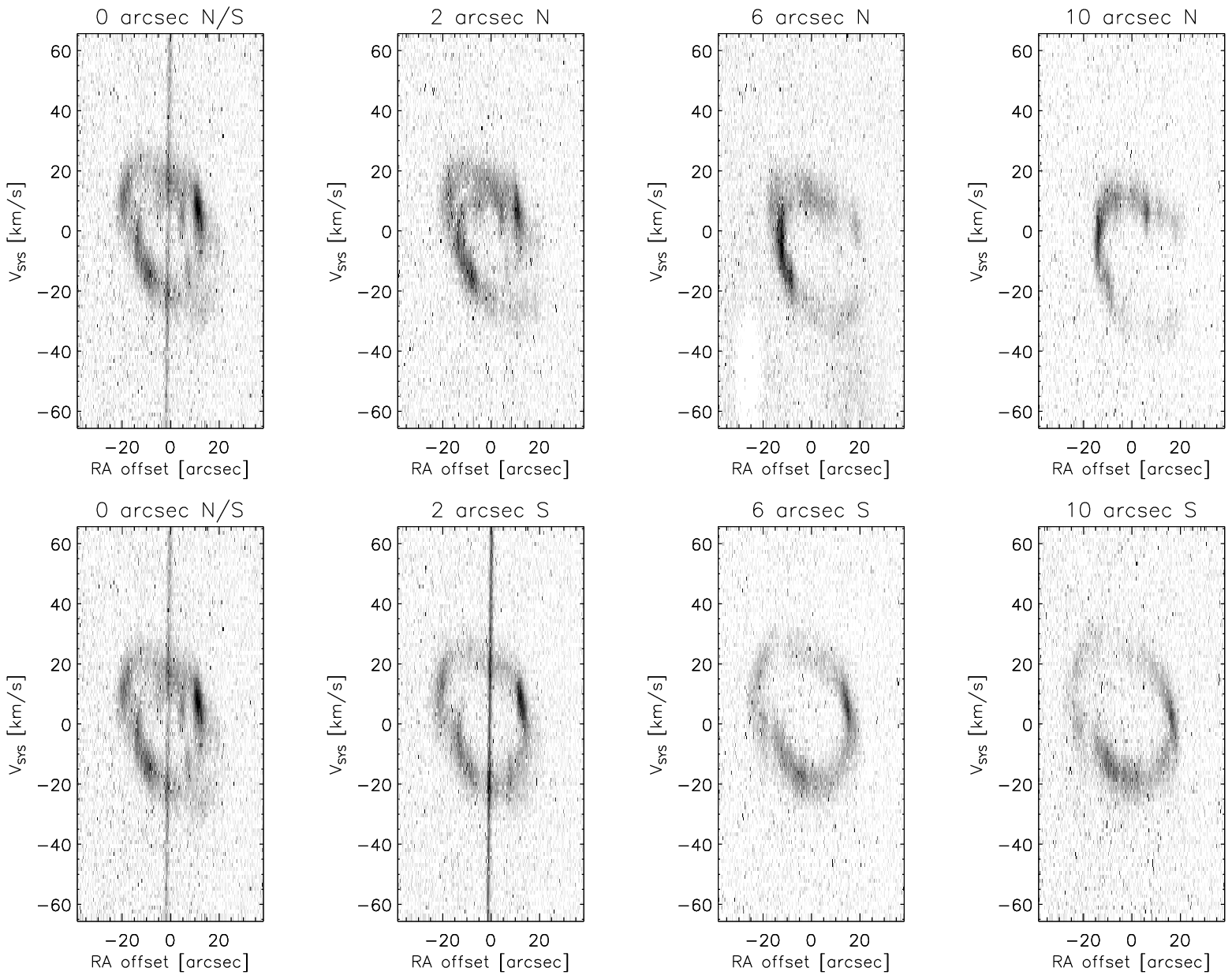}

\caption{Phoenix position-velocity images of NGC 2346 obtained
at the slit positions illustrated in Fig.\ 3. Top panels: images
obtained as the slit was stepped northward. Bottom panels: images
obtained as the slit was stepped southward.
The image in the leftmost panels in each series was obtained 
with the slit centered
on the waist of the nebula. The vertical band across the
images at offsets of $0''$ and $2''$S is continuum emission
from the binary companion to the central star (see Bond,
these proceedings).}

\end{figure}

In Fig.\ 3 we display an H$_2$ image of NGC 2346 obtained with the NOAO
Cryogenic Optical Bench (Kastner et  al.\ 1996) and we illustrate
the slit positions used for Phoenix spectroscopic observations.
Phoenix spectral images of NGC 2346 obtained at these positions
are presented in Fig.\ 4. These 
images demonstrate that the H$_2$ emission from the bipolar NGC 2346 
forms rings or ellipses in position-velocity space, an observation that 
reinforces our prior conclusion that ring-like planetaries which display 
H$_2$ are bipolar in structure (Kastner et  al.\ 1994, 1996).

The position-velocity
ellipse represented in the spectral image obtained with
the slit at $0''$ offset (leftmost panels) is noticably tilted, with the
largest redshifts found $\sim15''$ to the east and the
largest blueshifts $\sim15''$ to the west of the central
star. It is apparent from Fig.\ 3 that this tilt is due to
the orientation of the Phoenix slit with respect to the object. That
is, to the east of the star the slit takes in portions of
the rearward-facing (redshifted) south polar lobe of the
nebula, whereas to the west the slit samples portions of the
forward-facing (blueshifted) north polar lobe.

Furthermore, the position-velocity ellipses in Fig.\ 4
contain two distinct kinematic components: a central ring
associated with lower velocity material in the nebular waist 
and a pair of rings associated with higher velocity material 
in the bipolar outflow lobes.
The central ring is centered at the systemic velocity of the 
nebula and
is most apparent in the spectral images obtained near the
position of the central star (i.e., in the four lefthand
panels).  The southern ring 
is primarily redshifted (righthand bottom panels) 
while the northern ring (righthand top panels) is primarily
blueshifted. 
All three rings are present in the images
obtained nearest the position of the central star (lefthand
panels), whereas the images obtained further from the
central star display emission from only a portion of the central ring and
one of the outer rings. Hence
Figs.\ 3 and 4 indicate that the H$_2$ emission from the
nebula's waist produces the inner position-velocity ring,
while the outer rings arise from H$_2$ emission from the
polar lobes. Because of the tilt of the slit with respect to
the waist of the nebula, a given slit position samples both
the waist region and one or both polar lobes, resulting in a
superposition of these kinematic features in a given
spectral image.

In summary, the Phoenix spectral images of NGC 2346 provide
strong evidence for distinct kinematic components in this
nebula. These components consist of an equatorial ring or
disk which is expanding at relatively low velocity ($\sim15$
km s$^{-1}$ projected along our line of sight; Fig.\ 4,
leftmost panels) and polar lobes that are expanding at
larger velocities (Fig.\ 4, rightmost panels). Put differently, the
equatorial confinement that is apparent in the morphology of
this classical bipolar PN has a direct kinematic
counterpart. It is tempting, therefore, to conclude that the
pinched waist of NGC 2346 has its roots in processes which
we are now beginning to explore in objects such as RAFGL
2688.

\acknowledgments
J.H.K. acknowledges support from a JPL Infrared Space Observatory grant to
RIT. LeeAnn Henn (MIT) reduced many of the Phoenix spectral images used in
this study.

\end{document}